\documentclass{article}
\usepackage{spconf,amsmath,graphicx}

\usepackage{multirow}
\usepackage{cite}
\usepackage{setspace}

\title{CUHK-EE Voice Cloning System for ICASSP 2021 M2VoC Challenge}

\name{Daxin Tan, Hingpang Huang, Guangyan Zhang, Tan Lee}
\address{Department of Electronic Engineering, The Chinese University of Hong Kong, Hong Kong}

\begin{document}

\maketitle

\begin{abstract}
This paper presents the CUHK-EE voice cloning system for ICASSP 2021 M2VoC challenge. The challenge provides two Mandarin speech corpora: the AIShell-3 corpus of 218 speakers with noise and reverberation and the MST corpus including high-quality speech of one male and one female speakers. 100 and 5 utterances of 3 target speakers in different voice and style are provided in track 1 and 2 respectively, and the participants are required to synthesize speech in target speaker's voice and style. We take part in the track 1 and carry out voice cloning based on 100 utterances of target speakers. An end-to-end voicing cloning system is developed to accomplish the task, which includes: 1. a text and speech front-end module with the help of forced alignment, 2. an acoustic model combining Tacotron2 and DurIAN to predict melspectrogram, 3. a Hifigan vocoder for waveform generation. Our system comprises three stages: multi-speaker training stage, target speaker adaption stage and target speaker synthesis stage. Our team is identified as T17. The subjective evaluation results provided by the challenge organizer demonstrate the effectiveness of our system. Audio samples are available at our demo page.\footnote{https://daxintan-cuhk.github.io/CUHK-EE-system-M2VoC-challenge/}
\end{abstract}
\vspace{-0.5em}

\vspace{-1em}
\section{Introduction}
\vspace{-0.5em}
Text-to-speech (TTS) refers to the technology of generating natural speech from given text. Conventional approaches to TTS, e.g., unit selection \cite{hunt1996unit, pollet2017unit}, statistical parametric speech synthesis (SPSS) \cite{black2007statistical, koriyama2019statistical}, have been overtaken largely by deep neural network (DNN) based speech synthesis models in recent years \cite{shen2018natural,yu2019durian,ren2020fastspeech}. Given a reasonable amount of training data, end-to-end TTS systems are capable of producing speech with superior quality, nearly indistinguishable from human speech. 

A typical DNN based TTS system comprises an acoustic model and a vocoder. The former predicts spectrogram-like representations from text and the latter generates waveform from this representation. In Tacotron2\cite{shen2018natural}, the acoustic model is built with an autoregressive neural network. It learns the alignment between text and speech implicitly with location-sensitive attention mechanism. The DurIAN system \cite{yu2019durian} also uses the autoregressive model design and incorporates an explicit duration model. Fastspeech2 \cite{ren2020fastspeech} represents a non-autoregressive design of speech generation based on self-attention structure. As for the vocoder, a number of prominent neural models have been developed, e.g., Wavenet\cite{oord2016wavenet}, Waveglow\cite{prenger2019waveglow}, Melgan\cite{kumar2019melgan} and Hifigan\cite{kong2020hifi}. 

A key research issue in neural TTS is the control of speaker voice and style of synthesized speech. To produce speech with the voice and style of a target speaker is a challenging problem, especially when the utterances from the target speaker are limited. This problem is referred to as ``voice cloning''. The ICASSP 2021 M2VoC challenge (Multi-Speaker Multi-Style Voice Cloning Challenge)\cite{xie2021multi} is focused on investigating and evaluating state-of-the-art solutions to the problem of ``voice cloning''.

There are two common approaches to tackle with this problem: speaker encoding and speaker adaption. In speaker encoding method, speaker embedding is extracted from target speaker utterances via a speaker verification system, and speech are generated by conditioning on this speaker embedding and text\cite{jia2018transfer, cooper2020zero}. In speaker adaption method, a multi-speaker system is first established, and then the utterances of target speaker are used to adapt and fine-tune this multi-speaker system \cite{chen2018sample, hu2019neural}. Besides the voice characteristic, style can also be captured and transferred by style-related representation\cite{skerry2018towards, wang2018style, klimkov2019fine, tan2020fine}, which helps to incorporate the style of target speaker into synthesized speech.

Our model is a DNN based TTS system and mainly adopt the ``speaker adaption'' method. The rest of paper is organized as follows: the task is described in section 2. Our system is described in section 3. The evaluation results are discussed in section 4. The conclusions are given in section 5.

\vspace{-1em}
\section{The task in M2VoC challenge}
\vspace{-0.5em}
In M2VoC challenge, two Mandarin speech corpora are released for model development: the AIShell-3 corpus of 218 speakers with noise and reverberation and the MST corpus including high-quality speech of one male and one female speakers. 100 and 5 utterances of 3 target speakers in different voice and style (chat, game, story telling) are provided in track 1 and 2 respectively, and the participants are required to synthesize speech in target speaker's voice and style. Each track includes two subtracks, a and b. Participants can only use released data in subtrack a while use of external public data are allowed in subtrack b. We take part in the track 1. We develop our system only with the data released by organizer, and submit the synthesized audio in both track 1a and track 1b to see the influence of external data.

\vspace{-1em}
\section{System description}
\begin{figure}[h]
  \centering
  \includegraphics[width=\linewidth, trim=50 50 50 50]{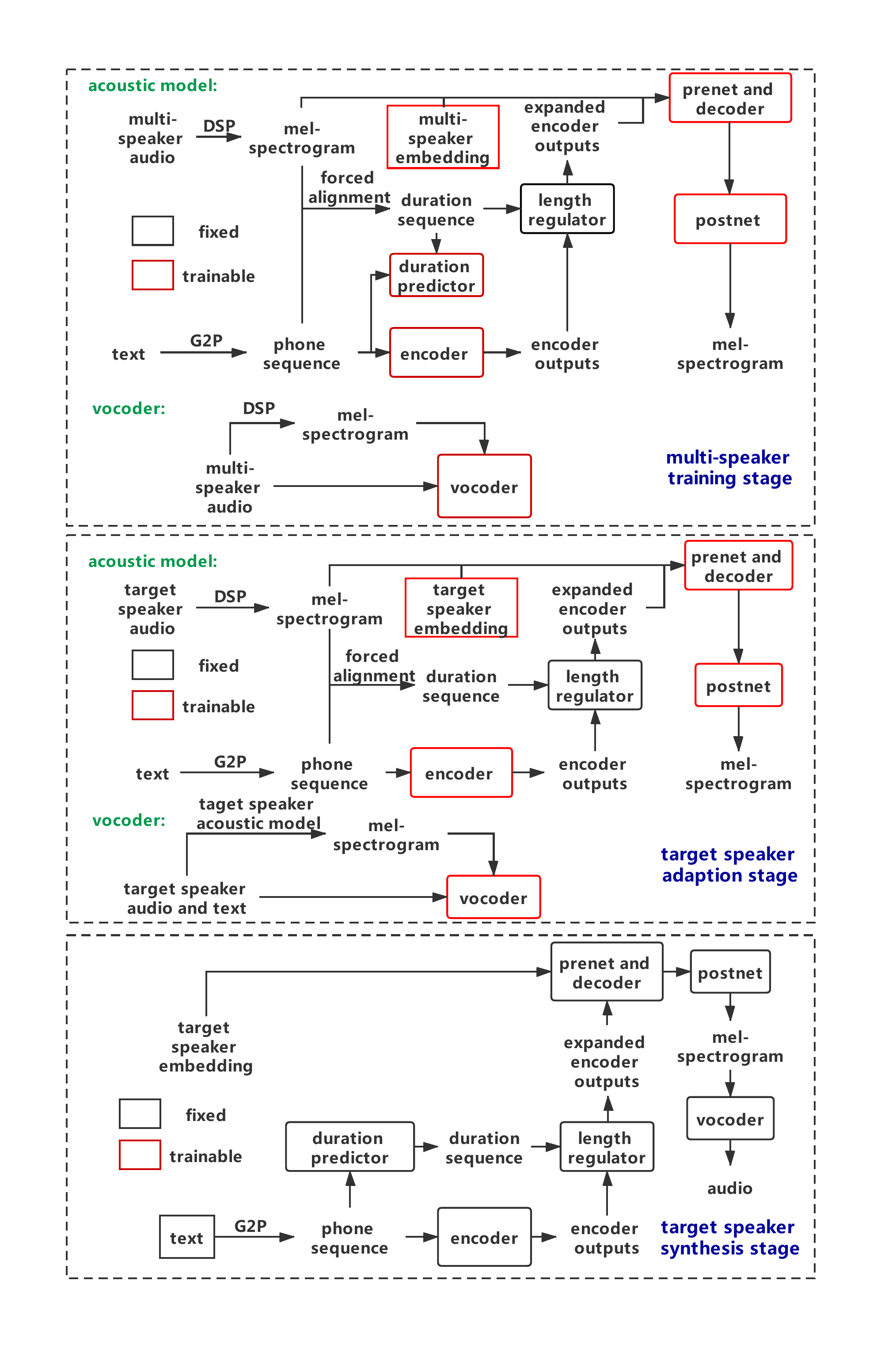}
  \caption{Overview of our system.}
  \label{fig:overview of system}
  \vspace{-1.5em}
\end{figure}

\vspace{-0.5em}
The overview of our system is shown in figure \ref{fig:overview of system}. Our system consists of three stages: multi-speaker training stage, target speaker adaption stage and target speaker synthesis stage. Our system is made up of three modules: text and speech front-end module, acoustic model and vocoder. 

\vspace{-1em}
\subsection{Text and speech front-end module}
\vspace{-0.5em}
Though texts and pinyin are both provided in the training corpus, we just make use of the texts and neglect the provided pinyin. The Chinese characters in texts are first converted to pinyin with 5 tones by a grapheme-to-phoneme (G2P) toolkit named pypinyin. Then HMM based Montreal Forced Aligner (MFA)\cite{mcauliffe2017montreal} is utilized to carry out the forced alignment. Specifically, the pinyin is first converted to global phone by a pre-trained G2P module and then a pre-trained acoustic model based on the same global phone set is used to carry out the forced alignment. In this way, the phone sequence of texts and the phone-level duration sequence are derived. It should be noted that, this phone sequence contains not only phone in global phone set but also ``silence'' and ``short pause'' symbols introduced in the speech. For the speech, we first convert all the audio to 22050 Hz and 16-bit PCM and normalize the loudness by Sox. Then short-time Fourier transform and 80-dimensional mel scale are applied to derive the mel-spectrogram. The window size and hop size are 1024 and 256 respectively and the mel filters span from 0 up to 8000 Hz, which follows the setting of Tacotron2. The duration sequence in time is then converted to the duration sequence in frame number. 

In the training and adaption stage, the ground-truth duration sequence is used for the training of the following acoustic model. As no ground-truth duration is known in advance in the synthesis stage, we train a duration predictor where ground-truth duration sequence and phone sequence serve as the training pair. The duration predictor consists of a bidirectional 512-dimensional LSTM and a linear layer while the loss is the MSE of duration in the log domain. It should be noted that, no prosodic boundary label is introduced in the phone sequence in the training stage, as the duration and short pause pattern in speech has already reflected the prosodic pattern in our viewpoint. In the synthesis stage, where only texts are provided without duration information, the texts are converted to pinyin then to global phone sequence by pypinyin and G2P in MFA respectively. Then punctuation like comma is converted to the ``short pause'' symbol and inserted into the phone sequence. Then this phone sequence is used to predict the duration sequence by the duration predictor.

\vspace{-1em}
\subsection{Acoustic model}
\vspace{-0.5em}
The acoustic model predicts mel-spectrogram from the phone sequence and duration sequence with assigned speaker embedding. Our acoustic model is a combination of Tacotron2 and DurIAN, which introduces duration information explicitly into Tacotron2. Our acoustic model consists of encoder, decoder, prenet, postnet, length regulator and speaker encoder. Specifically, the phone sequence is first processed by encoder, which consists of three convolutional layers and one bidirectional LSTM, to derive the encoder outputs in phone level. Then length regulator expands the encoder outputs to frame level according to the duration of each phone in frame number. The position embedding, representing the relative position of current frame in current phone by the linear interpolation between 0 and 1, is concatenated to encoder outputs of each step respectively. Then a speaker embedding obtained from speaker ID via speaker encoder, a trainable look up table, is concatenated to encoder outputs in all steps. These expanded encoder outputs and the ground-truth mel-spectrogram are used for decoder processing in training and adaption stage. For each step, the ground-truth frame of last step after prenet and the encoder output of last step are processed by a unidirectional LSTM to derive the context vector, then context vector and encoder output of current step are processed by another unidirectional LSTM and linear layer to obtain the mel-spectrogram of current step. A postnet module is applied to this mel-spectrogram to derive fine structure. The mel-spectrogram before and after postnet are both compared with the ground-truth mel-spectrogram to calculate MSE loss. This loss is used to optimize the encoder, decoder, prenet, postnet and speaker encoder in the acoustic model. In synthesis stage, the predicted duration sequence, phone sequence and speaker embedding are all used to predict the mel-spectrogram via acoustic model in an autoregressive manner.

Our preliminary experiments indicates that the acoustic model trained on corpus with more speakers but lower quality will results in low-quality synthesized speech while speaker similarity is not improved. Based on this observation, in the multi-speaker training stage, we just make use of MST 2-speaker high-quality speech corpus and discard the AIShell-3 218-speaker low-quaity speech corpus. In the target speaker adaption stage, only 100 utterances of the target speaker are used for the adaption of acoustic model. All the parameters in the acoustic model are fine-tuned with 3000 steps. 

\vspace{-0.5em}
\subsection{Vocoder}
\vspace{-0.5em}
The vocoder is used to generate the waveform from the mel-spectrogram. We utilize the Hifigan\cite{kong2020hifi} vocoder due to its high quality and fast generation speed. A GAN structre is adopted in the vocoder, where the generator is a fully convolutional network with multi-receptive field fusion while two discriminators, i.e., multi-period discriminator and multi-scale discriminator, are utilized to distinguish the ground-truth waveform and the waveform synthesized from mel-spectrogram. A pre-trained Hifigan vocoder is used in the multi-speaker training stage. In the target speaker adaption stage, after the acoustic model has been adapted, the ground-truth waveform and the mel-spectrogram generated from the adapted acoustic model serve as the training pair for vocoder adaption. 3000 steps of adaption is adopted in our experiments. The adapted vocoder is used in the synthesis stage.

\vspace{-0.5em}
\section{Results}
\vspace{-0.5em}
The official evaluation results of our system are presented in this section. In track 1, three sets of texts are provided by the organizer, including style set, common set and intelligibility set, each containing 100 sentences of one of three target speakers. The submitted speech are synthesized based on these three sets of texts. Only data released in the challenge can be used in the track 1a but external data can be used in the track 1b. For track 1, the final score is the average of mean opinion score (MOS) evaluation on speech quality, speaker similarity and style similarity. The evaluation is carried out by two rounds. All teams are involved in the first round and the top-scoring teams are involved in the second round. As our teams are not involved in the second round, the following analysis is based on the statistics of first round. In track 1, 19 systems (18 participating teams and one natural speech) and 23 systems (22 participating teams and one natural speech) are evaluated for track 1a and track 1b respectively. Our system is marked as T17 and the natural speech is TAR. 

\begin{figure}[h]
  \centering
  \includegraphics[width=\linewidth, trim=0 30 30 20]{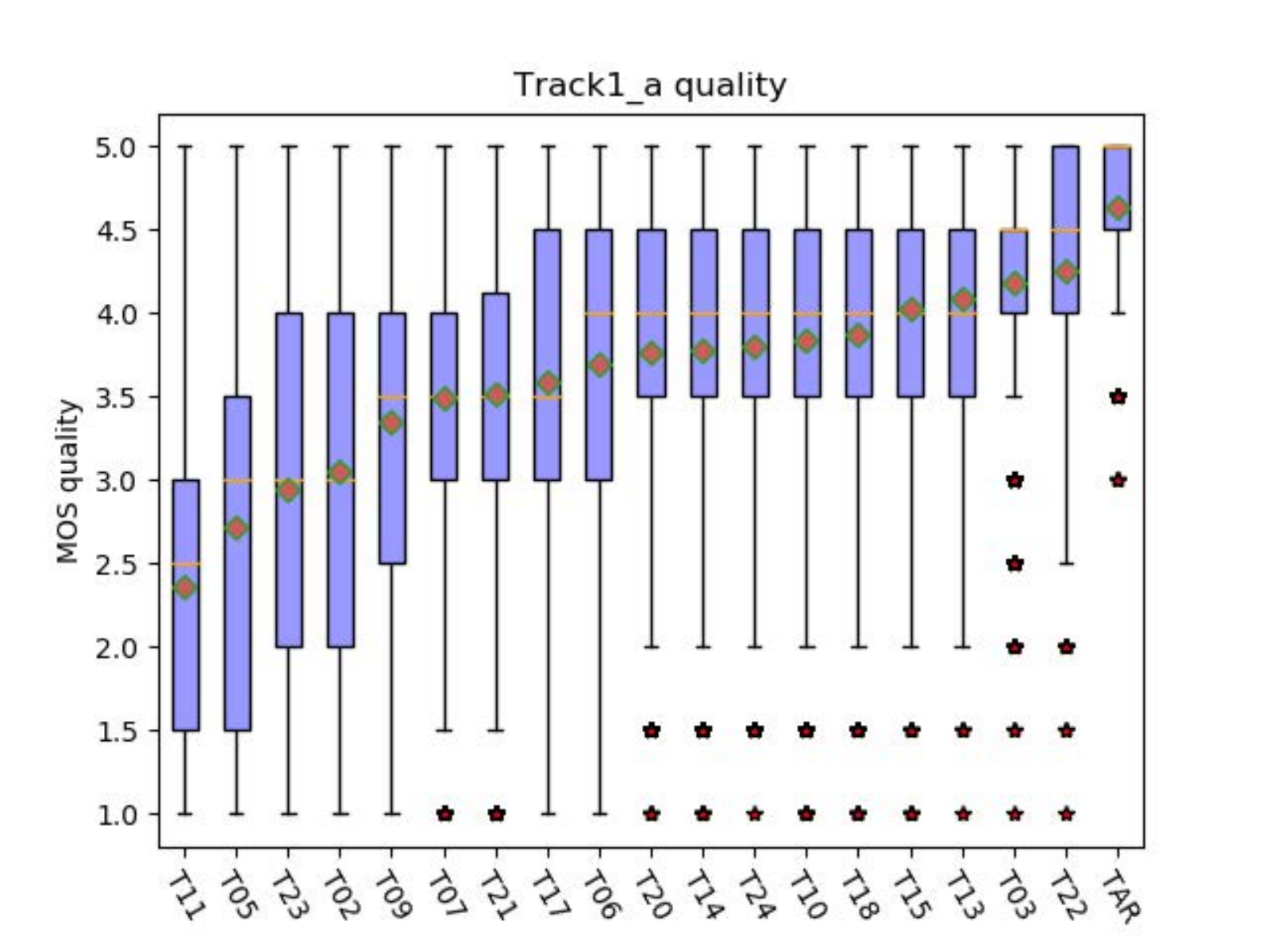}
  \caption{Boxplots of MOS of speech quality test in track 1a. TAR is natural speech and T17 is our system.}
  \label{fig:boxplot_track1a_quality}
\end{figure}

\vspace{-0.5em}
\begin{figure}[h]
  \centering
  \includegraphics[width=\linewidth, trim=0 30 30 25]{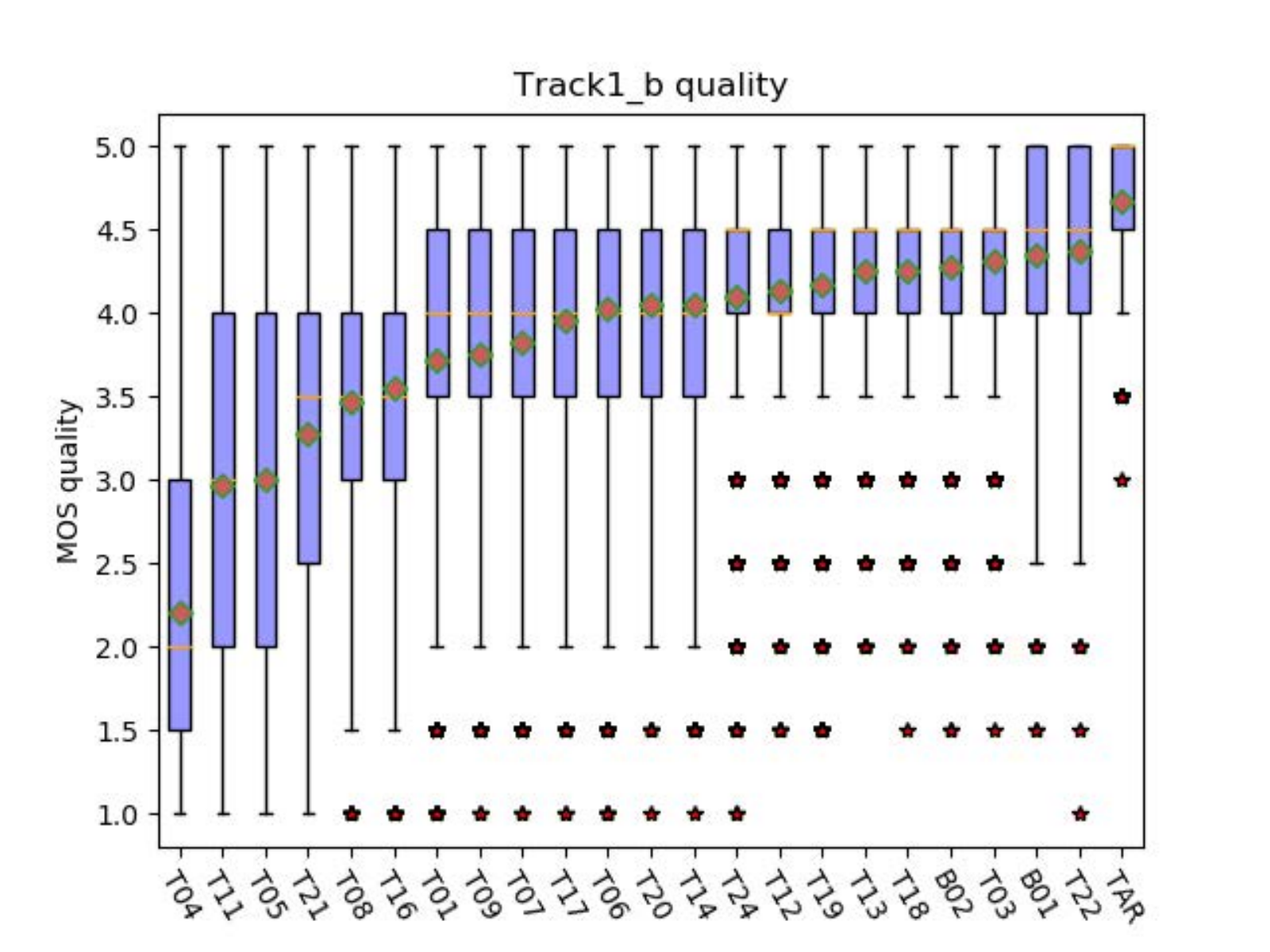}
  \caption{Boxplots of MOS of speech quality test in track 1b. TAR is natural speech and T17 is our system.}
  \label{fig:boxplot_track1b_quality}
  \vspace{-1em}
\end{figure}

\vspace{-0.5em}
\subsection{Speech quality test}
\vspace{-0.5em}
Figure \ref{fig:boxplot_track1a_quality} and \ref{fig:boxplot_track1b_quality} show the boxplots of MOS of each system on the speech quality in track 1a and track 1b respectively. The MOS score is on a scale of 1 (completely unnatural) to 5 (completely natural). In track 1a, the MOS of natural speech, 1st system and our system are 4.6410±0.0387, 4.2570±0.0414 and 3.5845±0.0563 respectively. Our team is ranked as 11 among the 18 submitted system. In track 1b, the MOS of natural speech, 1st system and our system are 4.6660±0.0359, 4.3780±0.0362, 3.9630±0.0468 respectively. Our team is ranked as 13 among the 22 submitted system.

\begin{figure}[h]
  \centering
  \includegraphics[width=\linewidth, trim=0 30 30 25]{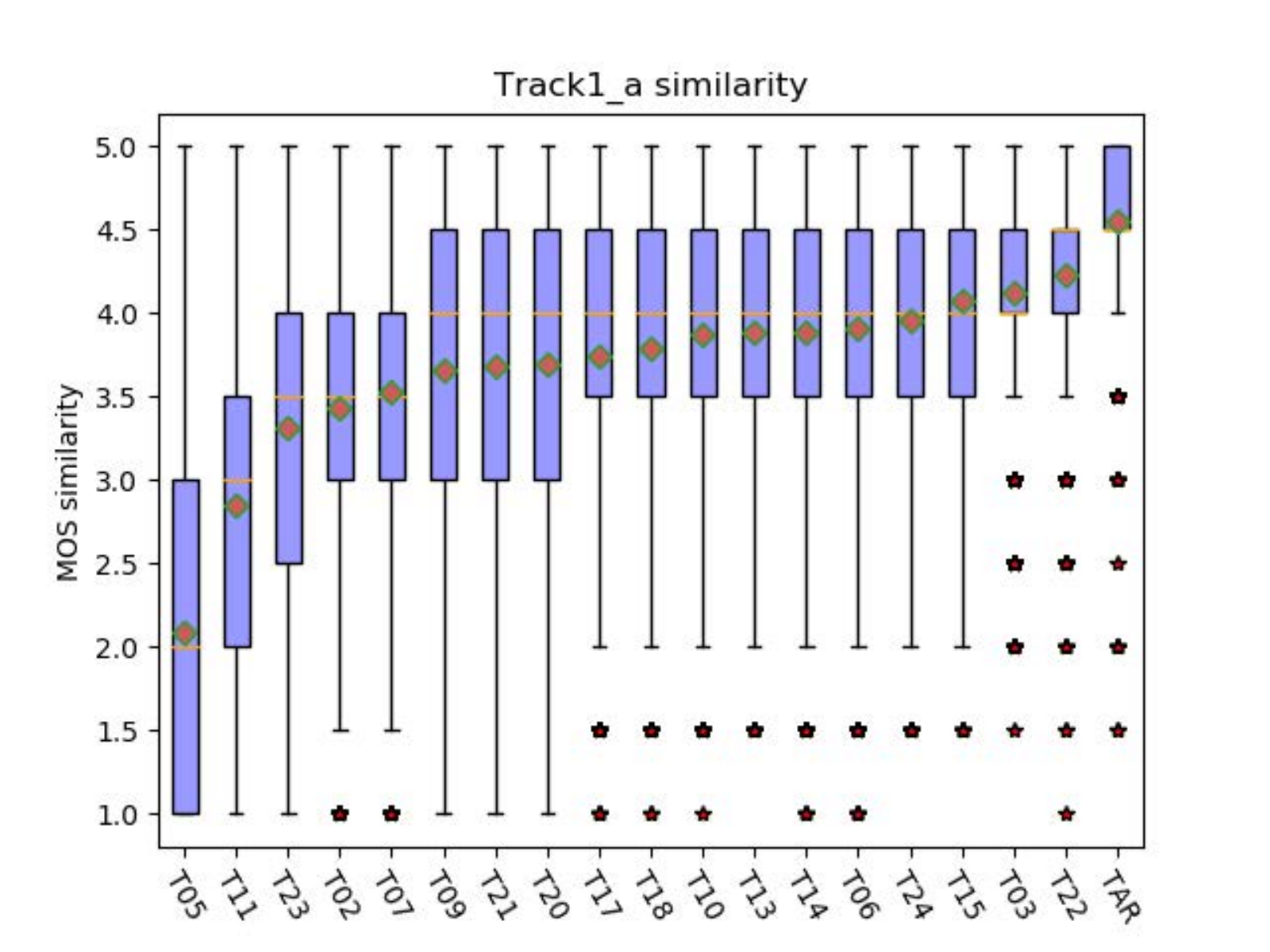}
  \caption{Boxplots of MOS of speaker similarity test in track 1a. TAR is natural speech and T17 is our system.}
  \label{fig:boxplot_track1a_similarity}
\end{figure}

\begin{figure}[h]
  \centering
  \includegraphics[width=\linewidth, trim=0 30 30 25]{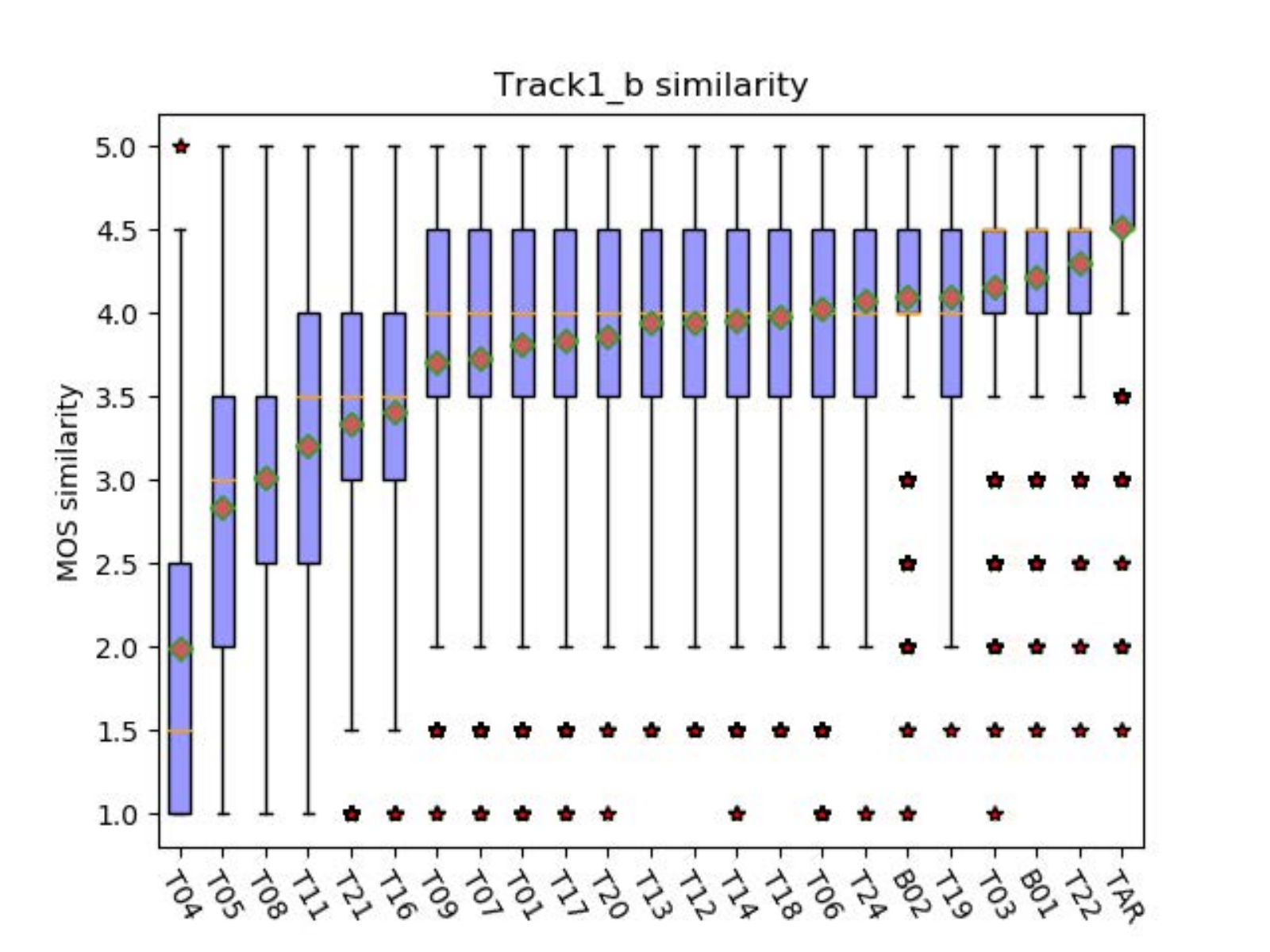}
  \caption{Boxplots of MOS of speaker similarity test in track 1b. TAR is natural speech and T17 is our system.}
  \label{fig:boxplot_track1b_similarity}
  \vspace{-1em}
\end{figure}

\vspace{-0.5em}
\subsection{Speaker similarity test}
\vspace{-0.5em}
Figure \ref{fig:boxplot_track1a_similarity} and \ref{fig:boxplot_track1b_similarity} show the boxplots of MOS of each system on the speaker similarity in track 1a and track 1b respectively. The MOS score is on a scale of 1 (Sounds like a totally different person) to 5 (Sounds like exactly the same person). In track 1a, the MOS of natural speech, 1st system and our system are 4.5500±0.0503, 4.2280±0.0371, 3.7470±0.0499 respectively. Our team is ranked as 10 among the 18 submitted system. In track 1b, the MOS of natural speech, 1st system and our system are 4.5180±0.0500, 4.2970±0.0349 and 3.8365±0.0499 respectively. Our team is ranked as 13 among the 22 submitted system.

\begin{figure}[h]
  \centering
  \includegraphics[width=\linewidth, trim=0 30 30 25]{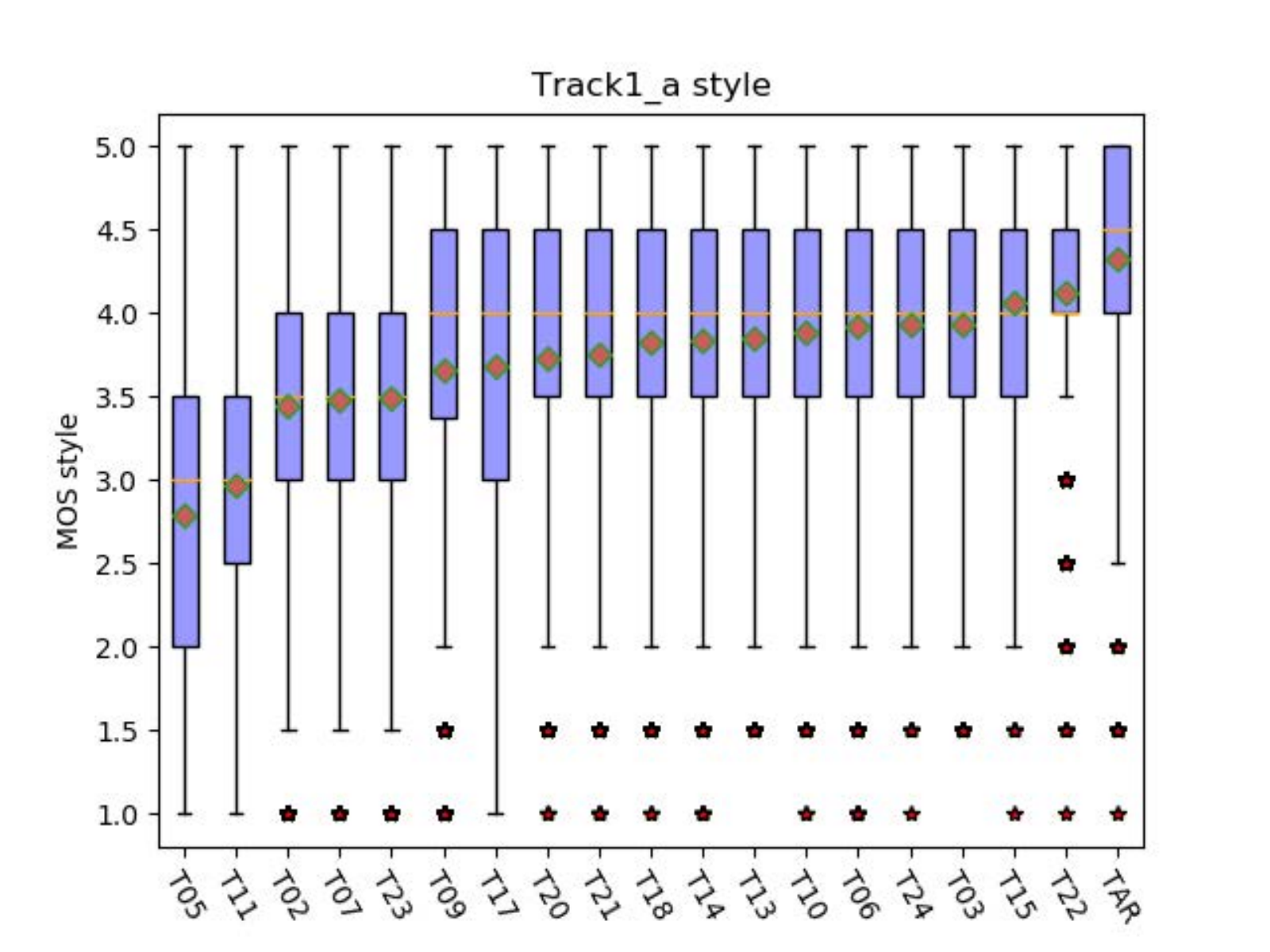}
  \caption{Boxplots of MOS of style similarity test in track 1a. TAR is natural speech and T17 is our system.}
  \label{fig:boxplot_track1a_style}
\end{figure}

\begin{figure}[h]
  \centering
  \includegraphics[width=\linewidth, trim=0 30 30 25]{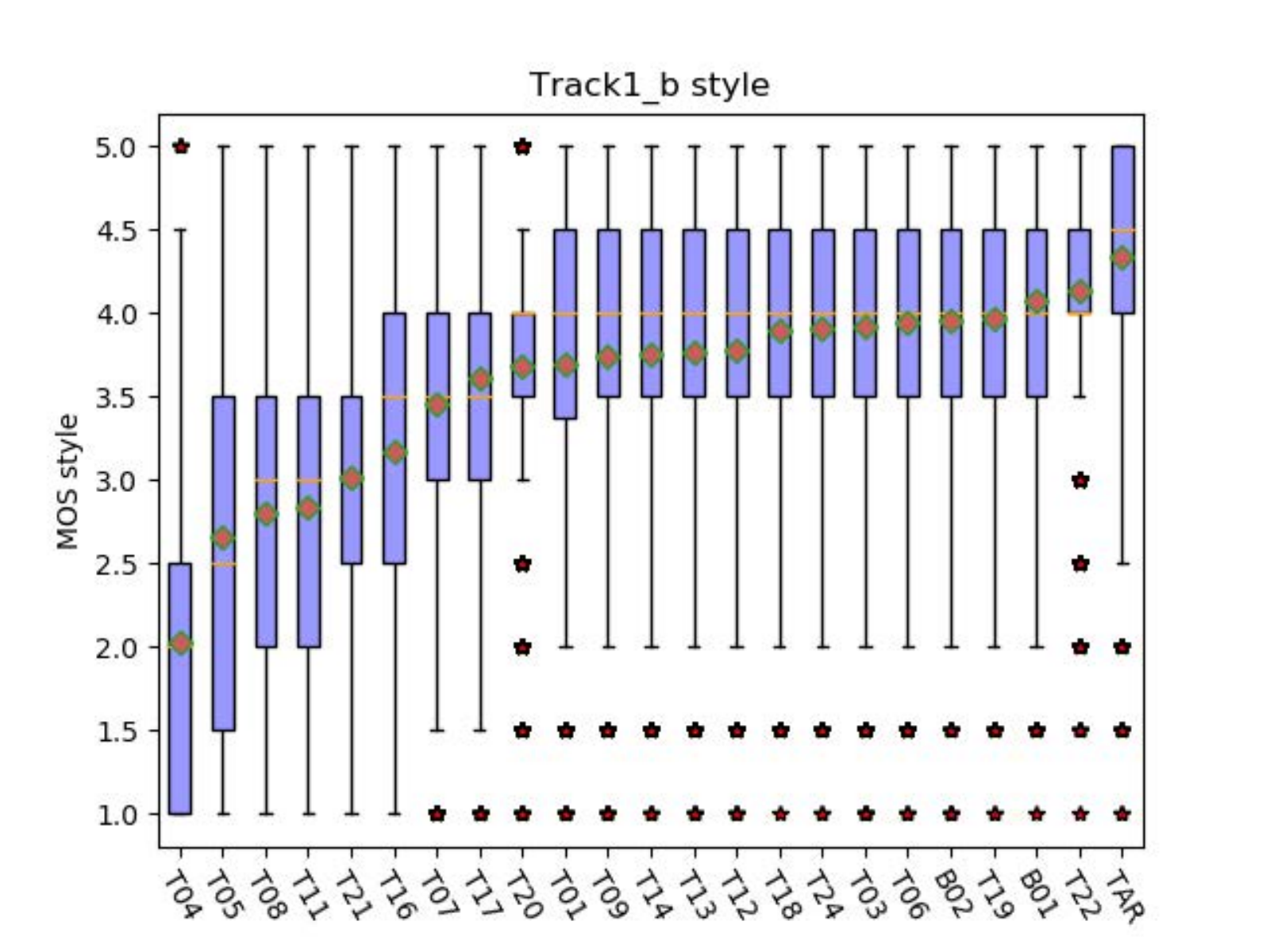}
  \caption{Boxplots of MOS of style similarity test in track 1b. TAR is natural speech and T17 is our system.}
  \label{fig:boxplot_track1b_style}
  \vspace{-1em}
\end{figure}

\vspace{-1em}
\subsection{Style similarity test}
\vspace{-0.5em}
Figure \ref{fig:boxplot_track1a_style} and \ref{fig:boxplot_track1b_style} show the boxplots of MOS of each system on the style similarity in track 1a and track 1b respectively. The MOS score is on a scale of 1 (sounds like a totally different style) to 5 (sounds like exactly the same style).
In track 1a, the MOS of natural speech, 1st system and our system are 4.3230±0.0338, 4.1210±0.0328 and 3.6840±0.0410 respectively. Our team is ranked as 12 among the 18 submitted system. In track 1b, the MOS of natural speech, 1st system and our system are 4.3410±0.0357, 4.1377±0.0333 and 3.6163±0.0417 respectively. Our team is ranked as 15 among the 22 submitted system.

\vspace{-1.5em}
\section{Conclusion}
\vspace{-1em}
This paper presents the details of our submitted system and results of ICASSP 2021 M2VoC challenge. An end-to-end voice cloning system is developed based on acoustic model of combination of Tacotron2 and DurIAN framework, followed by Hifigan vocoder. Three stages are involved in our system to carry out voice cloning. The subjective evaluation indicates that our system has performance of average level and have much room for improvement. In the future, we will explore more in the following aspects: the explicit use of prosodic boundary label,  the adaption of duration model, methods to encode the phone and tone in pinyin and exploration towards better acoustic model and vocoder.

\vfill\pagebreak


\begin{spacing}{0.8}
\bibliographystyle{IEEEbib}
\bibliography{mybib}

\begin{thebibliography}{10}

\bibitem{hunt1996unit}
Andrew~J Hunt and Alan~W Black,
\newblock ``Unit selection in a concatenative speech synthesis system using a
  large speech database,''
\newblock in {\em 1996 IEEE International Conference on Acoustics, Speech, and
  Signal Processing Conference Proceedings}. IEEE, 1996, vol.~1, pp. 373--376.

\bibitem{pollet2017unit}
Vincent Pollet, Enrico Zovato, Sufian Irhimeh, and Pier~Domenico Batzu,
\newblock ``Unit selection with hierarchical cascaded long short term memory
  bidirectional recurrent neural nets.,''
\newblock in {\em INTERSPEECH}, 2017, pp. 3966--3970.

\bibitem{black2007statistical}
Alan~W Black, Heiga Zen, and Keiichi Tokuda,
\newblock ``Statistical parametric speech synthesis,''
\newblock in {\em 2007 IEEE International Conference on Acoustics, Speech and
  Signal Processing-ICASSP'07}. IEEE, 2007, vol.~4, pp. IV--1229.

\bibitem{koriyama2019statistical}
Tomoki Koriyama and Takao Kobayashi,
\newblock ``Statistical parametric speech synthesis using deep gaussian
  processes,''
\newblock {\em IEEE/ACM Transactions on Audio, Speech, and Language
  Processing}, vol. 27, no. 5, pp. 948--959, 2019.

\bibitem{shen2018natural}
Jonathan Shen, Ruoming Pang, Ron~J Weiss, Mike Schuster, Navdeep Jaitly,
  Zongheng Yang, Zhifeng Chen, Yu~Zhang, Yuxuan Wang, Rj~Skerrv-Ryan, et~al.,
\newblock ``Natural tts synthesis by conditioning wavenet on mel spectrogram
  predictions,''
\newblock in {\em 2018 IEEE International Conference on Acoustics, Speech and
  Signal Processing (ICASSP)}. IEEE, 2018, pp. 4779--4783.

\bibitem{yu2019durian}
Chengzhu Yu, Heng Lu, Na~Hu, Meng Yu, Chao Weng, Kun Xu, Peng Liu, Deyi Tuo,
  Shiyin Kang, Guangzhi Lei, et~al.,
\newblock ``Durian: Duration informed attention network for multimodal
  synthesis,''
\newblock {\em arXiv preprint arXiv:1909.01700}, 2019.

\bibitem{ren2020fastspeech}
Yi~Ren, Chenxu Hu, Tao Qin, Sheng Zhao, Zhou Zhao, and Tie-Yan Liu,
\newblock ``Fastspeech 2: Fast and high-quality end-to-end text-to-speech,''
\newblock {\em arXiv preprint arXiv:2006.04558}, 2020.

\bibitem{oord2016wavenet}
Aaron van~den Oord, Sander Dieleman, Heiga Zen, Karen Simonyan, Oriol Vinyals,
  Alex Graves, Nal Kalchbrenner, Andrew Senior, and Koray Kavukcuoglu,
\newblock ``Wavenet: A generative model for raw audio,''
\newblock {\em arXiv preprint arXiv:1609.03499}, 2016.

\bibitem{prenger2019waveglow}
Ryan Prenger, Rafael Valle, and Bryan Catanzaro,
\newblock ``Waveglow: A flow-based generative network for speech synthesis,''
\newblock in {\em ICASSP 2019-2019 IEEE International Conference on Acoustics,
  Speech and Signal Processing (ICASSP)}. IEEE, 2019, pp. 3617--3621.

\bibitem{kumar2019melgan}
Kundan Kumar, Rithesh Kumar, Thibault de~Boissiere, Lucas Gestin, Wei~Zhen
  Teoh, Jose Sotelo, Alexandre de~Br{\'e}bisson, Yoshua Bengio, and Aaron
  Courville,
\newblock ``Melgan: Generative adversarial networks for conditional waveform
  synthesis,''
\newblock {\em arXiv preprint arXiv:1910.06711}, 2019.

\bibitem{kong2020hifi}
Jungil Kong, Jaehyeon Kim, and Jaekyoung Bae,
\newblock ``Hifi-gan: Generative adversarial networks for efficient and high
  fidelity speech synthesis,''
\newblock {\em arXiv preprint arXiv:2010.05646}, 2020.

\bibitem{xie2021multi}
Qicong Xie, Xiaohai Tian, Guanghou Liu, Kun Song, Lei Xie, Zhiyong Wu, Hai Li,
  Song Shi, Haizhou Li, Fen Hong, et~al.,
\newblock ``The multi-speaker multi-style voice cloning challenge 2021,''
\newblock in {\em ICASSP 2021-2021 IEEE International Conference on Acoustics,
  Speech and Signal Processing (ICASSP)}. IEEE, 2021, pp. 8613--8617.

\bibitem{jia2018transfer}
Ye~Jia, Yu~Zhang, Ron~J Weiss, Quan Wang, Jonathan Shen, Fei Ren, Zhifeng Chen,
  Patrick Nguyen, Ruoming Pang, Ignacio~Lopez Moreno, et~al.,
\newblock ``Transfer learning from speaker verification to multispeaker
  text-to-speech synthesis,''
\newblock {\em arXiv preprint arXiv:1806.04558}, 2018.

\bibitem{cooper2020zero}
Erica Cooper, Cheng-I Lai, Yusuke Yasuda, Fuming Fang, Xin Wang, Nanxin Chen,
  and Junichi Yamagishi,
\newblock ``Zero-shot multi-speaker text-to-speech with state-of-the-art neural
  speaker embeddings,''
\newblock in {\em ICASSP 2020-2020 IEEE International Conference on Acoustics,
  Speech and Signal Processing (ICASSP)}. IEEE, 2020, pp. 6184--6188.

\bibitem{chen2018sample}
Yutian Chen, Yannis Assael, Brendan Shillingford, David Budden, Scott Reed,
  Heiga Zen, Quan Wang, Luis~C Cobo, Andrew Trask, Ben Laurie, et~al.,
\newblock ``Sample efficient adaptive text-to-speech,''
\newblock {\em arXiv preprint arXiv:1809.10460}, 2018.

\bibitem{hu2019neural}
Qiong Hu, Erik Marchi, David Winarsky, Yannis Stylianou, Devang Naik, and
  Sachin Kajarekar,
\newblock ``Neural text-to-speech adaptation from low quality public
  recordings,''
\newblock in {\em Speech Synthesis Workshop}, 2019, vol.~10.

\bibitem{skerry2018towards}
RJ~Skerry-Ryan, Eric Battenberg, Ying Xiao, Yuxuan Wang, Daisy Stanton, Joel
  Shor, Ron Weiss, Rob Clark, and Rif~A Saurous,
\newblock ``Towards end-to-end prosody transfer for expressive speech synthesis
  with tacotron,''
\newblock in {\em international conference on machine learning}. PMLR, 2018,
  pp. 4693--4702.

\bibitem{wang2018style}
Yuxuan Wang, Daisy Stanton, Yu~Zhang, RJ-Skerry Ryan, Eric Battenberg, Joel
  Shor, Ying Xiao, Ye~Jia, Fei Ren, and Rif~A Saurous,
\newblock ``Style tokens: Unsupervised style modeling, control and transfer in
  end-to-end speech synthesis,''
\newblock in {\em International Conference on Machine Learning}. PMLR, 2018,
  pp. 5180--5189.

\bibitem{klimkov2019fine}
Viacheslav Klimkov, Srikanth Ronanki, Jonas Rohnke, and Thomas Drugman,
\newblock ``Fine-grained robust prosody transfer for single-speaker neural
  text-to-speech,''
\newblock {\em arXiv preprint arXiv:1907.02479}, 2019.

\bibitem{tan2020fine}
Daxin Tan and Tan Lee,
\newblock ``Fine-grained style modeling, transfer and prediction in
  text-to-speech synthesis via phone-level content-style disentanglement,''
\newblock {\em arXiv preprint arXiv:2011.03943}, 2020.

\bibitem{mcauliffe2017montreal}
Michael McAuliffe, Michaela Socolof, Sarah Mihuc, Michael Wagner, and Morgan
  Sonderegger,
\newblock ``Montreal forced aligner: Trainable text-speech alignment using
  kaldi.,''
\newblock in {\em Interspeech}, 2017, vol. 2017, pp. 498--502.

\end{thebibliography}
\end{spacing}

\end{document}